\documentclass[12pt]{iopart}
\begin{document}

\input{psfig} 

\letter{Divergence of Dipole Sums and the
  Nature of Non-Lorentzian Exponentially Narrow Resonances in
  One-Dimensional Periodic Arrays of Nanospheres}
 
\author{Vadim A. Markel
\footnote[3]{e-mail: vmarkel@mail.med.upenn.edu}} 

\address{Departments of Radiology and Bioengineering, \\
  University of Pennsylvania, Philadelphia, PA 19104}

\begin{abstract}
  Origin and properties of non-Lorentzian spectral lines in linear
  chains of nanospheres are discussed. The lines are shown to be
  super-exponentially narrow with the characteristic width
  $\propto\exp[-C(h/a)^3]$ where $C$ is a numerical constant, $h$ the
  spacing between the nanospheres in the chain and $a$ the sphere
  radius. The fine structure of these spectral lines is also
  investigated.
\end{abstract}

\submitto{\JPB}
%\maketitle
\vspace{2cm}

One-dimensional periodic chains (ODPC) of metallic nanospheres have
attracted significant recent attention due to their unusual optical
properties.  Although the general theoretical framework for analyzing
electromagnetic interactions in ODPC has been built a decade
ago~\cite{markel_93_1}, the recent dramatic advances in
nanofabrication have reinvigorated the interest in ODPC, which, in
turn, has led to several new results of high experimental relevancy.
In particular, radiatively non-decaying surface plasmons (SPs) in ODPC
with possible applications to building novel lasers were discussed in
Ref.~\cite{burin_04_1}; unusual shifts of plasmon resonance
frequencies were found in Ref.~\cite{zhao_03_1} and a dramatic narrowing of
SP spectral lines was found in Ref.~\cite{zou_04_1,zou_04_2} in finite
chains of moderate length. In this letter I show that two of these
phenomena (unusual shifts and narrowing of SP spectral lines) are
directly related to a logarithmic divergence of dipole sums
(electromagnetic eigenvalues) - a theoretical interpretation that has
not been given so far. The SPs that can be excited as a result of this
divergence posses highly unusual properties. In particular, the
resonance line-shapes are essentially non-Lorentzian and are
characterized by a vanishing integral weight. This is in a sharp
contrast to spectral line broadening or narrowing due to change in the
decay rate. Another consequence which has not been previously noticed
is that each narrow resonance is paired with an even more narrow
spectral hole. An interesting new property discussed below is that the
narrow collective SP resonances can be excited in ODPCs even when the
distance between neighboring spheres is much larger than the sphere
diameters. However, the resonances become so narrow in this case that,
unless one takes special care, it is extremely unlikely to notice them
in any numerical or experimental investigation.

\begin{figure}
\centerline{\psfig{file=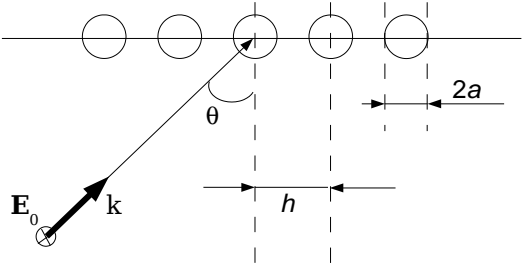,width=8.5cm,bbllx=0bp,bblly=375bp,bburx=600bp,bbury=675bp,clip=t}}
\caption{Sketch of the ODPC excitation by a linearly polarized plane
  wave.}
\label{fig:sketch}
\end{figure}

We start with a brief summary of the underlying physics. The geometry
of a ODPC excited by a plane wave of the form ${\bf E}_0\exp(i{\bf
  k}\cdot{\bf r})$ is illustrated in Fig.~\ref{fig:sketch}. Here the
polarization vector ${\bf E}_0$ is perpendicular to the chain, the
incidence angle is denoted by $\theta$, lattice spacing by $h$ and the
sphere radius by $a$. Each sphere is assumed to be polarizable and
characterized by the dipole polarizability $\alpha$. We work in the
approximation introduced by Doyle~\cite{doyle_89_1} in which each
sphere is treated as an elementary dipole located at its center but is
characterized by non-quasistatic polarizability $\alpha$ which is
calculated from the coefficient $a_1$ of the Mie
theory~\cite{bohren_book_83}:

\begin{equation}
\label{alpha}
\alpha = {{3i} \over {2k^3}} {{m\psi_1(mka)\psi_1^{\prime}(ka) -
    \psi_1(ka)\psi_1^{\prime}(mka)} \over {m\psi_1(mka)
    \xi_1^{\prime}(ka) - \xi_1(ka)\psi_1^{\prime}(mka)}} \ ,
\end{equation}

\noindent
where $\psi_1(x)$ and $\xi_1(x)$ are the Riccati-Bessel functions,
$m=\sqrt{\epsilon}$ is the complex refractive index of the spheres and
$k=\omega/c=2\pi/\lambda$ is the wave number of the incident wave. The
above approximation allows one to include spheres which are not small
compared to the incident wavelength $\lambda$ while staying within the
purely dipole theory. The higher multipole interactions of the
spheres, as well as the input of higher multipoles to the optical
cross sections, are ignored in this approximation. Note that the
polarizability $\alpha$ defined by (\ref{alpha}) is the {\em exact}
dipole polarizability with respect to excitation by a plane wave, but
not by secondary waves scattered by spheres in the chain.  However,
the dipole approximation was shown to be very accurate when $h$ is the
order of or larger than $2a$ (which is the case discussed below) by
direct comparison with a converged T-matrix solution~\cite{zou_04_1}.
In general, it is known that short-range multipole interactions of
orders higher than the first (dipole) do not play a significant role
for transverse electromagnetic excitations of finite or infinite
linear arrays of interacting spheres even when the spheres are in
close proximity~\cite{ruppin_89_1,mazets_00_1}.  Physically, one can
argue that the short-range interaction is not important for transverse
excitations because it does not lead to an electric current along the
chain (in a sharp contrast to the longitudinal excitations which are
not discussed in the letter).

\begin{figure}
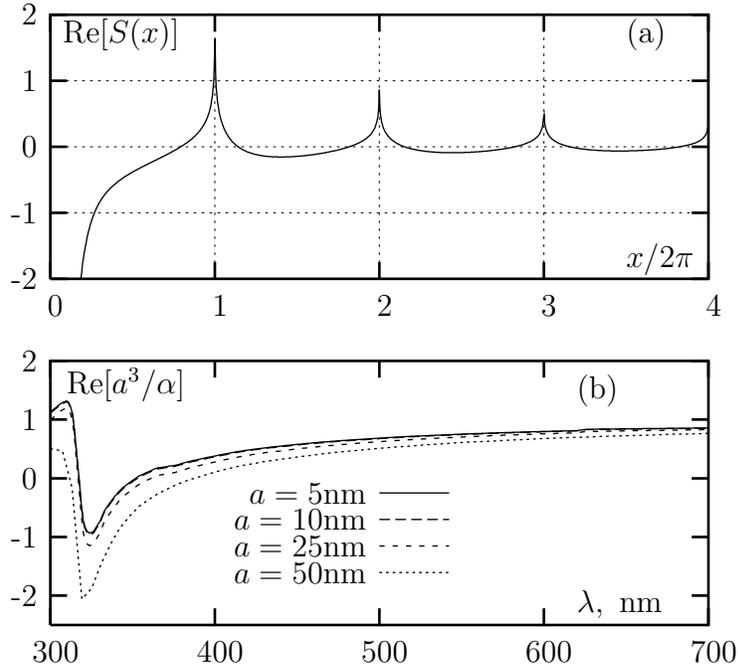

\centerline{\input{fig2a.tex}}
\centerline{\input{fig2b.tex}}
\caption{(a) Function ${\rm Re}S(x)$ for $\theta=0$.  (b) ${\rm
    Re}(a^3/\alpha)$ calculated for silver for different values of
  $a$; the curves $a={\rm 5nm}$ and $a={\rm 10nm}$ correspond to the
  quasi-static limit and are indistinguishable.}
\label{fig:x_delta}
\end{figure}

In the approximation formulated above, each sphere is characterized by
a dipole moment which, in the case of geometry shown in
Fig.~\ref{fig:sketch}, is collinear with the polarization vector
${\bf E}_0$ and has the amplitude $d_n$, $n$ being the index which
labels spheres in the chain. The amplitudes $d_n$ are coupled to the
external field and to each other by the coupled-dipole
equation~\cite{markel_93_1}

\begin{equation}
\label{CDE}
d_n = \alpha\left[E_0\exp(ikn\sin\theta) + \sum_{n^{\prime}\neq n}
  W_{n-n^{\prime}}(kh) d_{n^{\prime}} \right] \ .
\end{equation}

\noindent
Here $W_{n-n^{\prime}}$ is the dipole interaction term given by

\begin{equation}
W_n(x) = k^3\left( {x^{-1} \over{\vert n\vert}} +
  {ix^{-2} \over{\vert n \vert^2}} - {x^{-3} \over{\vert
      n \vert^3}} \right) \exp(ix\vert n \vert) \ .
\label{W_nn}
\end{equation}

\noindent
The coupled-dipole equation (\ref{CDE}) is easily solved to yield 

\begin{equation}
\label{d_n}
d_n = {{ a^3E_0\exp(ikhn\sin\theta)} \over {a^3/\alpha - (ka)^3 S(kh)}}
\ ,
\end{equation}

\noindent
where the dimensionless quantity $(ka)^3 S(kh)$ is the dipole sum that
characterizes excitation of the SP with the wave number
$q=k\sin\theta$. The function $S(x)$ is given by

\begin{equation}
S(x) = 2 \sum_{n>0} \left({ 1 \over {xn}} + {i \over {(xn)^2}} - {1
    \over {(xn)^3}} \right)
\exp(inx)\cos(nx\sin\theta)  \ .
\label{S_def}
\end{equation}

\noindent
It can be seen that the above series diverges logarithmically when $(1 \pm
\sin\theta)kh = 2\pi l$, $l$ being an integer.  It is convenient to
separate the sum into two parts: $S(x)=S_1(x) + S_2(x)$ where $S_1$ is
given by

\begin{equation}
\label{S_1_def}
S_1(x) = 2\sum_{n>0} {{\cos(nx)\cos(nx\sin\theta)} \over {nx}}
=-{1 \over {2x}}\ln\left[4(\cos x - \cos(x\sin\theta))^2\right] 
\end{equation}

\noindent
and diverges when $\cos x = \cos(x\sin\theta)$ while $S_2(x)$ is the
reminder of series (\ref{S_def}) and converges for all values of
parameters.  For simplicity, we will assume everywhere below normal
incidence ($\theta=0$) which was also the case considered in
Ref.~\cite{zou_04_1}. Then the divergence takes place when $h=\lambda
l$.

The specific extinction cross section per one sphere is given
by~\cite{markel_93_1}

\begin{equation}
\label{sigma_def}
\sigma_e = {\rm Im} {{4\pi k a^3} \over {a^3/\alpha - (ka)^3 S(kh)}} \
.
\end{equation}

\noindent
Optical resonances occur when the real part of the denominator in the
above expression vanishes. In Fig.~\ref{fig:x_delta}(a) we plot ${\rm
  Re}S(x)$ for $\theta=0$. The sharp peaks in the plot correspond to
the points where ${\rm Re}S$ diverges. In Fig.~\ref{fig:x_delta}(b) we
also plot ${\rm Re}(a^3/\alpha)$ calculated according to (\ref{alpha})
for silver.  Interpolated experimental dielectric function
from~\cite{palik_book_v1_85} was used in calculations. It can be seen
that in an isolated sphere, the SP (Frohlich) resonance takes place in
the interval ${\rm 350 nm}<\lambda < {\rm 380 nm}$, depending on the
value of $a$. The resonant wavelength is obtained from ${\rm
  Re}(a^3/\alpha)=0$. Above the Frohlich resonance (at smaller
wavelengths), the spectral variable ${\rm Re}(a^3/\alpha)$ becomes
negative. (Here we ignore the region $\lambda<{\rm 320 nm}$ where no
resonance excitation can take place due to the strong interband
absorption.) Therefore, in order to excite a SP in an interacting ODPC
in this spectral region, the variable ${\rm Re}S$ must be also
negative. As can be seen from Fig.~\ref{fig:x_delta}(a), this happens
for sufficiently small values of the parameter $x=kh$ and corresponds
to the conventional blue shift of the transverse electromagnetic
oscillations, which is well known and can be described by the
quasistatic interaction~\cite{markel_92_1}.  However, below the
Frohlich resonance (at larger wavelength), the spectral parameter
${\rm Re}(a^3/\alpha)$ is positive. Therefore, in order to excite
an SP in this spectral region, the variable ${\rm Re}S$ must be also
positive.  Obviously, this requirement is fulfilled near the points of
divergence of ${\rm Re}S$. Thus, if $h/\lambda$ is close to an
integer, the transverse collective oscillations of the chain are
shifted to the red from the Frohlich wavelength, contrary to the usual
case. Quite remarkably, the collective resonance can take place even
if $a<<h$ and $(ka)^3<<1$. Indeed, no matter how small $ka$ is, the
resonant condition can always be satisfied sufficiently close to the
point $\lambda = h/l$.  Below, we focus on the first of these
resonances, which corresponds to $\lambda \approx h$ and can be
experimentally observed in metal ODPCs in the visible and IR spectral
regions below the Frohlich frequency of an isolated sphere.

\begin{figure}
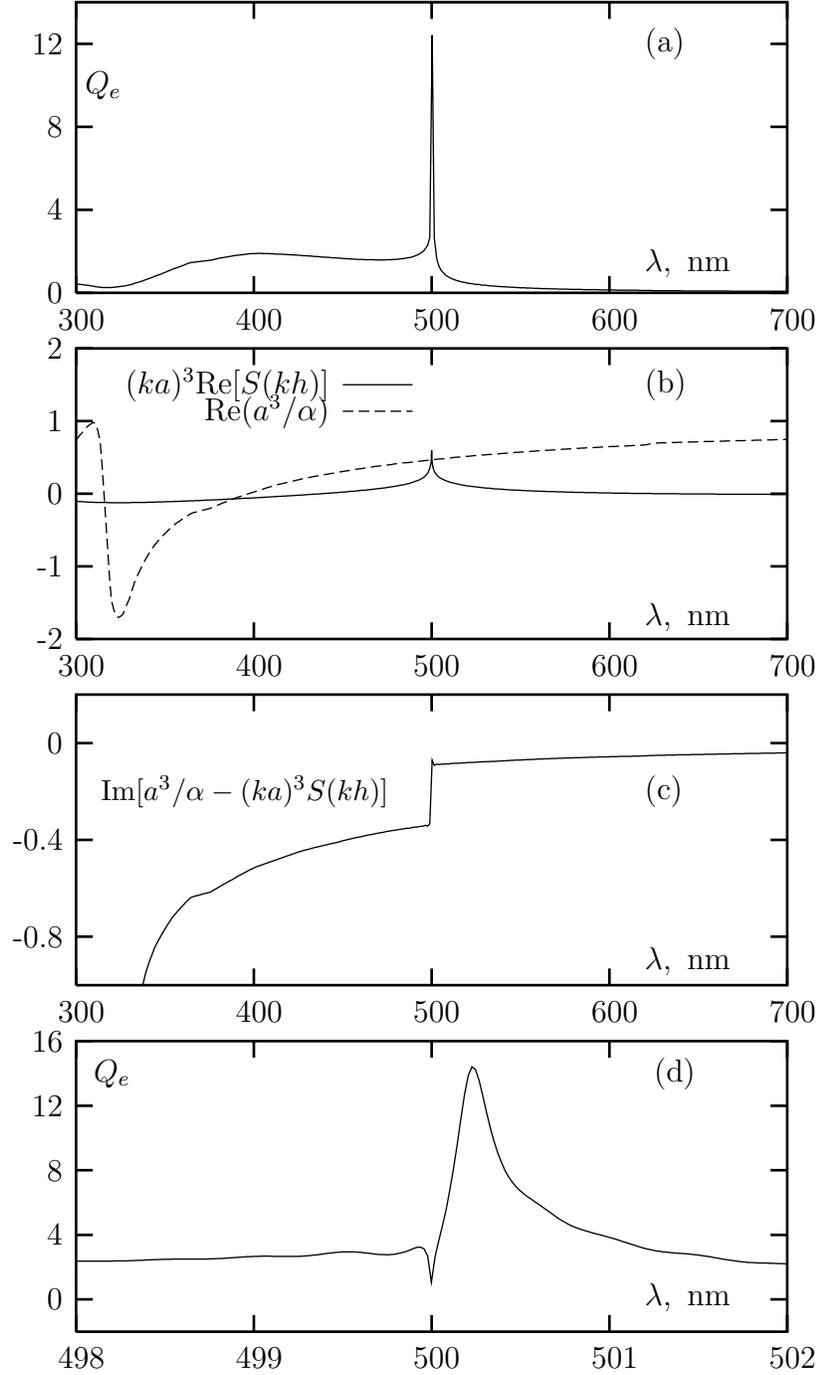

\centerline{\input{fig3a.tex}}
\centerline{\input{fig3b.tex}}
\centerline{\input{fig3c.tex}}
\centerline{\input{fig3d.tex}}
\caption{(a) Extinction efficiency $Q_e$ for $h=500{\rm nm}$ and
$a=50{\rm nm}$. (b) Illustration of cancellation of the real part of
the denominator in (\ref{sigma_def}). (c) Spectral dependence of the
imaginary part of the denominator. (d) The narrow resonance shown in
the panel (a) fully resolved.}
\label{fig:spectra}
\end{figure}

In Fig.~\ref{fig:spectra}(a) we plot the dimensionless extinction
efficiency $Q_e=\sigma_e/4\pi k a^3$ for the following values of
parameters: $a=50{\rm nm}$ and $h=500{\rm nm}$. The sharp resonance
corresponding to $\lambda\approx h$ is clearly visible. The curve is
very close to the one shown in Fig.~1(a) of Ref.~\cite{zou_04_1}. In
Fig.~\ref{fig:spectra}(b) the origin of the sharp resonance is
illustrated. Namely, we see that in a very narrow spectral interval
near $\lambda=500{\rm nm}$ real parts of the two terms in the
denominator of (\ref{sigma_def}) cancel each other. From the analysis
of Fig.~\ref{fig:spectra}(b), it is obvious that there should be, in
fact, two closely spaced narrow resonances separated by a spectral
hole. However, the first of the two resonances is suppressed due to
radiative losses. The imaginary part of the denominator of
(\ref{sigma_def}) is plotted in Fig.~\ref{fig:spectra}(c). As was
pointed out in Ref.~\cite{markel_93_1}, the imaginary part of $S(x)$
experiences a jump at $x=2\pi$. However, no exact cancellation of the
imaginary part of the denominator can take place.  For the geometry
considered here, it can be shown that ${\rm Im}[(ka)^3 S(kh)] >
-5(ka)^3/12$ while ${\rm Im}(a^3/\alpha) < -2(ka)^3/3$, so that the
imaginary part of the denominator is $\leq - (ka)^3/4$, with the
equality taking place for nonabsorbing materials with ${\rm
  Im}\epsilon=0$.  Finally, In Fig.~\ref{fig:spectra}(d) the sharp
resonance seen in Fig.~\ref{fig:spectra}(a) is completely resolved.
The narrow spectral hole located exactly at $\lambda=h$ can be also
seen in this figure.

Let us estimate the width and amplitude of the narrow resonances
occurring due to the divergence of $S(x)$. We define the width of the
resonance as the distance from the resonance wavelength $\lambda_r$,
determined from the condition ${\rm Re}[a^3/\alpha(\lambda_r) - (2\pi
a/\lambda_r)^3 S(2\pi h/\lambda_r)]=0$ to the center of the spectral
hole at $\lambda=h$; thus, $\Delta\lambda=\vert \lambda_r -
h\vert$. We estimate $\lambda_r$ assuming that the dominant
contribution to ${\rm Re}S$ comes from the logarithmically diverging
term (\ref{S_1_def}). We also assume that $\Delta\lambda<<2\pi$ and
expand the argument of the cosine in (\ref{S_1_def}) near the point
$x=kh=2\pi$, which leads to the following estimate:

\begin{equation}
\label{Delta_lambda}
\Delta\lambda \approx {h \over 2\pi} \exp\left[- {{C} \over
    {2(2\pi)^2}} \left({h \over a}\right)^3 \right] \ ,
\end{equation}

\noindent
where $C={\rm Re}[a^3/\alpha(\lambda=h)]$ is a constant of the order
of unity. For example, for $h=500{\rm nm}$, $C\approx 0.5$. Using
$a=50{\rm nm}$, we obtain from (\ref{Delta_lambda}) $\Delta\lambda
\approx 0.14{\rm nm}$ in agreement with Fig.~\ref{fig:spectra}(d).
Thus, the width of the resonance is completely determined by the
geometrical factors (the ratio $h/a$) and is not in any way controlled
by relaxation. The latter, however, influences amplitude of the
resonance. Indeed, the maximum value of $Q_e$ in the peak is given by
$1/{\rm Im}[a^3/\alpha - (ka)^3S(kh)]$. For the geometry considered
here, it can be verified that this value can not be greater than
$(h/\pi a)^2$, which is the limit for nonabsorbing material. However,
for strongly absorbing materials amplitude of the resonance can become
negligibly small.

Since the amplitude of the narrow resonances does not increase when
the width decreases super-exponentially, it is impossible to
effectively excite these resonances by a near-field probe. For
example, consider the case when a single sphere (say, $n=0$) is
excited by a near-field microscope tip of small aperture. Then the
coupled-dipole equation for the amplitudes $d_n$ can be solved by
Fourier transformation:

\begin{equation}
\label{d_n_Fur}
d_n = \int_{-\pi/h}^{\pi/h} {{a^3E_0\exp(iqhn)} \over {a^3/\alpha -
    (ka)^3 \tilde{S}(kh,qh)}} {h dq \over 2\pi} \ ,
\end{equation}

\noindent
where $\tilde{S}(kh,qh)$ is given by (\ref{S_def}) in which
$x\sin\theta$ in the argument of cosine must be formally substituted
by $qh$ and the variable $x$ in the reminder of the formula
substituted by $kh$. Function $\tilde{S}(kh,qh)$ diverges
logarithmically when $\cos(kh)=\cos(qh)$; in particular, if $kh=2\pi$
as in the examples considered above, the only point of divergence
within the integration interval is $q=0$. Similarly to resonances in
extinction spectra, this resonance is super-exponentially narrow in the
SP wavenumber $q$, and its input into the above integral is
negligible. We emphasize again that the resonances discussed here are
essentially non-Lorentzian, and the conditions for applicability of
the quasiparticle pole approximation, which under normal circumstances
would properly describe coupling of the near-field probe to SPs, are
severely violated.

\begin{figure}
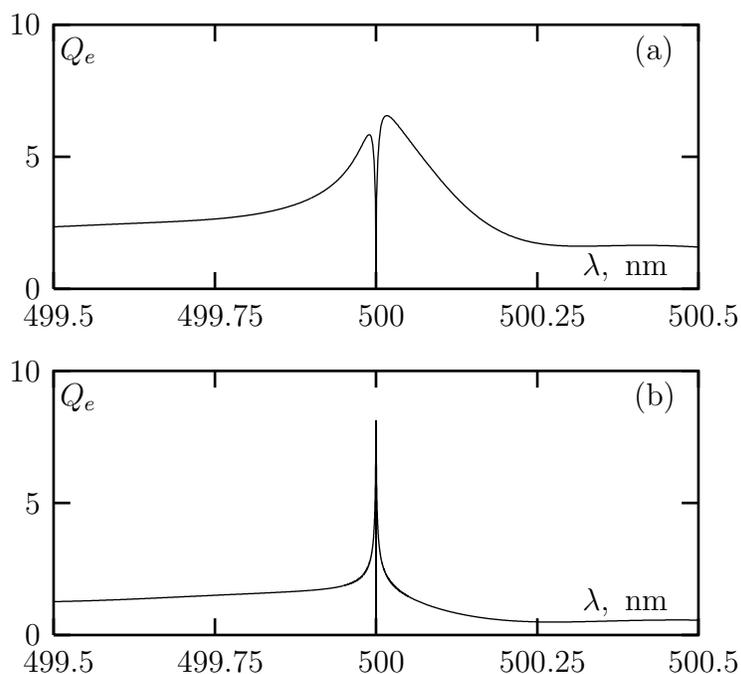

\centerline{\input{fig4a.tex}}
\centerline{\input{fig4b.tex}}
\caption{(a) Narrow resonances for $h=500{\rm nm}$ and $a=45{\rm nm}$
  (a) and $a=40{\rm nm}$ (b).}
\label{fig:narrow}
\end{figure}

The narrow resonance in Fig.~\ref{fig:spectra}(d) was obtained for
$h=500{\rm nm}$ and $a=50{\rm nm}$. We now show that the narrow
resonances also exist for smaller values of $a$ and larger ratios
$h/a$. To this end, we plot the extinction efficiency $Q_e$ for
$h=500{\rm nm}$ and $a=45{\rm nm}$ [Fig~\ref{fig:narrow}(a)] and
$a=40{\rm nm}$ [Fig~\ref{fig:narrow}(b)]. The dielectric function does
not vary noticeably over the narrow spectral range shown in
Fig.~\ref{fig:narrow} and was therefore taken to be constant,
$\epsilon=-8.5+0.76i$, which corresponds to the experimental value at
$\lambda=h=500{\rm nm}$ given in Ref.~\cite{palik_book_v1_85}. The
narrow non-Lorentzian resonances are well manifested in
Fig.~\ref{fig:narrow}. The central spectral hole is resolved to some
degree in Fig.~\ref{fig:narrow}(a) but is shown only as a vertical
line in Fig.~\ref{fig:narrow}(b).  Obviously, it is impossible to
resolve the spectral holes completely, since they do not have the
Lorentzian structure and are non-differentiable at the point
$\lambda=h$.  Note that the resonances shown in this figure are much
more symmetrical with respect to the point $\lambda=500{\rm nm}$ than
the one shown in Fig.~\ref{fig:spectra}(d). This is due to the fact
that with increasing the ratio $h/a$, the influence of radiative
losses on the shape of resonance lines decreases. For even larger
values of $h/a$, the resonances quickly become extremely narrow but do
not disappear completely, at least in chains of sufficient length.

It is interesting to consider the possibility of narrow resonances in
situations when the dipole approximation is not applicable, i.e., for
spheres in close proximity. It can be shown that the resonances
discussed in this letter do not disappear or get broadened when the
full multipole interaction is taken into account. Furthermore, the
spherical shape of the particles is also not fundamental because the
phenomenon discussed here originates due to long-range interaction in
ODPC while the higher multipole interaction is short-range. This
conclusion is in agreement with the numerical study of extinction
spectra of periodic chains of cylindrical disks~\cite{zou_04_2} which
were shown to have sharp resonances similar to those found in chains
of spheres.  Generalization to two-dimensional arrays of particles is
also possible. Two-dimensional periodically modulated structures have
also attracted significant recent attention, with the possible
application including random lasers~\cite{burin_04_1}, development of
novel chemical and biological sensors~\cite{zhao_03_1} and the study
of anomalous optical transmission through metal
films~\cite{darmanyan_04_1}.

To conclude this letter, we discuss several factors that contribute to
broadening of the spectral lines discussed above. The most important
factor is the finite length of a chain, since the divergence of the
dipole sums is logarithmic. As was mentioned above, narrow resonances
very close to those in infinite chains were obtained
in Ref.~\cite{zou_04_1} for $h/a=10$ and only 50 spheres; however,
observing more narrow resonances with $h/a>10$ will require a
substantially larger number of spheres. One possible solution to this
problem is to place the ODPC into a circular optical fiber.  Disorder
is another important factor. Numerical simulations in finite chains
(400 particles)~\cite{zou_04_2} revealed that random uncorrelated
displacements of particles with the amplitude of $\sim 0.1h$ do not
noticeably change the resonance lineshape. This is an expected result
for short-range disorder, i.e., the disorder with the correlation
length of one or few lattice spacings. However, disorder with
long-range correlations can result is much stronger changes in the
resonance lineshapes. Further, the account of nonlocality of the
dielectric response will not alter the nature of positive interference
(synchronism) which results in the logarithmic divergences, and is not
expected to broaden the narrow spectral lines.  The two physical
phenomena whose effects on the spectral lines discussed here are
difficult to predict are the nonlinearity of the optical response
(e.g., Kerr-type third-order nonlinearity) and quantum effects. These
effects will be the subject of future work.

%\bibliographystyle{prsty}
%\bibliography{abbrevplain,article,book}

\end{document}